\documentclass[12pt,singlespace,oneside]{article}
\usepackage{amsmath}
\usepackage{latexsym}

\input{tcilatex}

\begin{document}

\title{Real ``Units Imaginary''\linebreak\ in K\"{a}hler's Quantum Mechanics%
\thanks{%
Dedicated to Professor A. Rueda, in recognition for the impact he had on my
professional development.}\ }
\author{Jose G. Vargas\thanks{%
138 Promontory Rd., Columbia, SC 29209, USA. josegvargas@earthlink.net} \ }
\date{}
\maketitle

\begin{abstract}
Inspired by a similar, more general treatment by K\"{a}hler, we obtain the
spin operator by pulling to the Cartesian coordinate system the azimuthal
partial derivative of differential forms. At this point, no unit imaginary
enters the picture, regardless of whether those forms are over the real or
the complex field. Hence, the operator is to be viewed as a real operator.
Also a view of Lie differentiation as a pull-back emerges, thus avoiding
conceps such as flows of vector fields for its definition.

Enter Quantum Mechanics based on the K\"{a}hler calculus. Independently of
the unit imaginary in the phase factor, the proper values of the spin part
of angular momentum emerge as imaginary because of the idempotent defining
the ideal associated with cylindrical symmetry. Thus the unit imaginary has
to be introduced by hand as a factor in the angular momentum operator ---and
as a result also in its orbital part--- for it to have real proper values.
This is a concept of real operator opposite to that of the previous
paragraph. K\"{a}hler stops short of stating the antithesis in this pair of
concepts, both of them implicit in his work.

A solution to this antithesis lies in viewing units imaginary in those
idempotents as being the real quantities of square $-1$ in rotation
operators of real tangent Clifford algebra. In so doing, one expands the
calculus, and launches in principle a geometrization of quantum mechanics,
whether by design or not.
\end{abstract}

\section{Introduction}

In the standard treatment of angular momentum in the literature, the orbital
one comes first, through replacement with quantum mechanical operators for $%
\mathbf{r}$ and $\mathbf{p}$ in $\mathbf{r}\times \mathbf{p}$. One then
invokes internal degrees of freedom to attach spin to it, which is a little
bit awkward since spin has to do with rotations in configuration space, not
in some internal space. In addition, cylindrical symmetry is handled by in
the not too transparent way of ignoring part of what is conserved in the
spherical case. This is not too transparent since we have conservation for
all directions at the same time in the spherical case, but its manifestation
is curtailed by non-commutativity of operators for conservation of angular
momentum in different directions. Ideally, one should not resort to $\mathbf{%
r}\times \mathbf{p}$, which does not pertain to just cylindrical symmetry.
Finally, the unit imaginary makes explicit appearance through the $\mathbf{p}
$ operator in the standard treatment. This appearance is totally unnecessary
in K\"{a}hler's approach to quantum mechanics through his calculus, though
the unit imaginary still becomes necessary in the applications. We end these
considerations on angular momentum by quoting from E. K\"{a}hler \cite{K62}:

\begin{quotation}
``The spin of the electron will be interpreted as the necessity of
representing its state not by a wave function but by a wave differential
form'' (translated from the German original).
\end{quotation}

K\"{a}hler's treatment of angular momentum is the result of his extension of
the concept of Lie derivative from the ring of functions to the ring of
differential forms. But, in his case as in Cartan's, the concept of Lie
derivative of a differential form is a consequence of what it is for a
function ($0-$form), not a matter of an ad hoc definition. In fact, Cartan
did not even define what is now considered to be Lie differentiation of
differential forms, but did formulate a famous theorem, which can be used to
compute it.

In standard quantum mechanics, linear momentum and particles come to the
fore from the start. Thus, concepts like  linear momentum are foundational
in Dirac's approach. In K\"{a}hler'sK\"{a}hler, they certainly remain
relevant, but as concomitant or derived issues, not as foundational ones. In
fact, K\"{a}hler did not even directly deal with square-integrability, and
the position operator was not even mentioned in his work.

We obtain K\"{a}hler's expression for angular momentum starting which the
action of a partial derivative on a differential form. He obtained it in the
general framework of what Cartan called \cite{Cartan22} infinitesimal
transformations, which he cleverly transformed into partial derivatives \cite%
{K60}. But this is a distraction that obscures the humble nature of the Lie
derivative of a differential form. It is simply a pull-back to a different
coordinate system of a partial derivative of a differential form. This
simplicity is best appreciated when, searching the web, one realizes the
confusion that surrounds the concept of Lie differentiation.

Except for the difference just mentioned, our argument is totally K\"{a}%
hler's up to this point, the unit imaginary having not yet appeared. It is
at the point of obtaining solutions to equations with cylindrical symmetry
that he required the presence of something like the unit imaginary. But K%
\"{a}hler did not try to relate expressions without the unit imaginary to
expressions with it. In dealing with this issue, we find that the geometric
objects of square $-1$ that suggest themselves as alternatives for the unit
imaginary in the treatment of rotations are different in principle from
those in the idempotents that define related ideals. The solution to this
leads to what we consider the beginning of a process of geometrization of
quantum mechanics.

In sections 2 and 3, we shall simplify K\"{a}hler's treatment of spin in
order to target it directly without reference to the general subject of Lie
differentiation. Section 4 shows to non-experts how one deals with rotations
in Clifford algebra. This knowledge is used to infer in section 5 the case
by case replacement of the unit imaginary in real quantities in phase
shifts, rotations and energy and angular momentum operators.

\section{Lie differentiations of differential forms as pull-backs of their
partial differentiations}

We begin by reversing an argument by K\"{a}hler in 1960 \cite{K62}.
Basically, he started with an operator $\zeta ^{j}(x)\frac{\partial }{%
\partial x^{j}}$ and converted it into a partial derivative $\partial
/\partial y^{n}$ in some appropriately chosen coordinate system ($y^{j}$).
We are interested in a simpler problem. We have some partial derivative $%
\partial /\partial \phi $ acting on a differential form and we want to pull
that action to another coordinate system. Once we do that, we shall
specialize to when $\phi $ is the azimuthal coordinate in 3D-Euclidean space.

Give any (usually local) coordinate system ($z^{i}$) on a manifold, each $%
dz^{i}$ is an exact differential form whose evaluation on curves between
points $A$ and $B$ yields the curve-independent difference $%
z_{B}^{i}-z_{A}^{i}.$ We shall assume that (if it were not obvious from the
meaning of ) $\partial (dz^{m})/\partial z^{l}$ equals zero for all $l$ and $%
m$ independently of whether $l$ equals $m$ or not.

Let $\phi $ be a coordinate in some coordinate system ($y^{j}$), and let ($%
x^{i}$) be some overlapping coordinate system. As just stated, $\partial
\phi /\partial y^{l}=0$ and, therefore,

\begin{equation}
\frac{\partial (dx^{i})}{\partial \phi }=\frac{\partial }{\partial \phi }%
\frac{\partial x^{i}}{\partial y^{l}}dy^{l}=\frac{\partial }{\partial y^{l}}%
\left( \frac{\partial x^{i}}{\partial \phi }\right) dy^{l}=d\frac{\partial
x^{i}}{\partial \phi }=d\zeta ^{i},
\end{equation}%
where we have defined%
\begin{equation}
\zeta ^{i}\equiv \frac{\partial x^{i}}{\partial \phi },
\end{equation}%
and where we use throughout Einstein's summation convention over repeated
indices, one up and one down. Thus, for an arbitrary $1-$form, we have%
\begin{equation}
\frac{\partial (u_{i}dx^{i})}{\partial \phi }=u_{i},_{\phi
}dx^{i}+u_{i}d\zeta ^{i}=\zeta ^{j}\frac{\partial u_{i}}{\partial x^{j}}%
dx^{i}+d\zeta ^{i}u_{i}.
\end{equation}

Let $u$ denote now any differential form in the exterior or in the Clifford
algebras of differential forms built upon the n-dimensional module spanned
by the $dx^{i}.$ Those algebras are themselves modules of dimension $2^{n}$.
Hence, we can span $u$ as%
\begin{equation}
u=u_{\Lambda }dx^{\Lambda },\text{ \ \ \ \ \ \ \ \ \ \ \ \ }\Lambda
=1,2,...,2^{n}
\end{equation}%
where $dx^{\Lambda }$ represents the different elements of a basis of
differential forms in the $2^{n}$-dimensional module. For simplicity, each
element of the basis is taken to be of definite grade, going from $0$ to $n.$
Easy calculations show that, instead of (3), we would now have%
\begin{equation}
\frac{\partial u}{\partial \phi }=\zeta ^{j}\frac{\partial u_{\Lambda }}{%
\partial x^{j}}dx^{\Lambda }+d\zeta ^{i}\wedge e_{i}u,
\end{equation}%
where $e_{i}u$ is defined by%
\begin{equation}
u=\text{ }dx^{i}\wedge e_{i}u\text{ }+\text{ }u^{\prime },
\end{equation}%
and that no term in $u^{\prime }$ contains the factor $dx^{i}.$ If $u$ were
just a scalar function $f$, the second term in (5) would disappear, the only
non-vanishing $dx^{\Lambda }$ is the unity and $u_{\Lambda }$ is $f$ for
that $dx^{\Lambda }$.

Warning: Lie operators as defined by their action on differential forms do
not constitute a vector space. This can be seen as follows. Let the system ($%
y^{i}$) coincide with the system ($x^{i}$). Make $\phi $ be the last
coordinate. Then $\zeta ^{j}$=(0,0,...,1). Equation (5) reduces to%
\begin{equation}
\frac{\partial u}{\partial x^{n}}=\frac{\partial u_{\Lambda }}{\partial x^{n}%
}dx^{\Lambda }.
\end{equation}%
If we were dealing with a vector space, the Lie operator $\zeta ^{j}\partial
/\partial x^{j}$ acting on $u$ would yield only the first of the two terms
on the right hand side of (5).

In order to minimize overlooking these facts, we shall use the notation $%
\chi $ for the pull back of the operator $\frac{\partial }{\partial \phi }$
that acts on differential forms in the $x$ coordinate system. This addresses
a deficiency in notation of (5). But, we shall also avoid using the symbol $%
\frac{\partial u}{\partial x^{j}}$ for $\frac{\partial u_{\Lambda }}{%
\partial x^{j}}dx^{\Lambda }$, which is inimical to the result just obtained.

Assume $\phi $ were the azimuthal coordinate in 3-D Euclidean space. Then $%
\zeta ^{j}$=($-y,x,0$) and%
\begin{equation}
\frac{\partial u}{\partial \phi }=x\frac{\partial u}{\partial y}-y\frac{%
\partial u}{\partial x}-dy\wedge e_{1}u+dx\wedge e_{2}u,
\end{equation}%
where the subscripts $1$ and $2$ stand for the $x$ and $y$ coordinate
respectively. The terms $-dy\wedge e_{1}u+dx\wedge e_{2}u$ should not be
ignored. They constitute the action of the spin operator, as we shall show
further below.

\section{Spin for cylindrical symmetry}

We associate with $\chi $ a differential $1-$form $\tau $ $=\zeta _{i}dx^{i}$%
. For the azimuthal coordinate,%
\begin{equation}
\tau =-ydx+xdy,\text{ \ \ \ \ }d\tau =2dxdy.
\end{equation}

Define differential forms $u_{0}$ to $u_{3}$ that do not contain $dx$ and $%
dy $ by%
\begin{equation}
u=dxdyu_{0}+dxu_{1}+dyu_{2}+u_{3}.
\end{equation}%
One readily gets that%
\begin{equation}
d\tau u=-2u_{0}-dyu_{1}+dxu_{2}+dxdyu_{3},
\end{equation}%
and%
\begin{equation}
ud\tau =-2u_{0}+dyu_{1}-dxu_{2}+dxdyu_{3}
\end{equation}%
Hence%
\begin{equation}
\frac{1}{2}(d\tau u-ud\tau )=-dyu_{1}+dxu_{2}.
\end{equation}%
On the other hand%
\begin{equation}
-dy\wedge e_{1}u=-dy\wedge u_{1},\text{ \ \ \ \ }dx\wedge e_{2}u=dx\wedge
du_{2},
\end{equation}%
which then yields%
\begin{equation}
\frac{\partial u}{\partial \phi }=x\frac{\partial u}{\partial y}-y\frac{%
\partial u}{\partial x}+\frac{1}{2}d\tau u-\frac{1}{2}ud\tau .
\end{equation}

We define%
\begin{equation}
w_{1}=dydz,\text{ \ \ \ \ }w_{2}=dzdx,\text{ \ \ \ \ }w_{3}=dxdy,
\end{equation}%
and rewrite (15) as 
\begin{equation}
\chi _{3}=x^{1}\frac{\partial u}{\partial x^{2}}-x^{2}\frac{\partial u}{%
\partial x^{1}}+\frac{1}{2}w_{3}u-\frac{1}{2}uw_{3}.
\end{equation}

We have not used spherical symmetry, and we have not invoked the unit
imaginary for any purpose. It simply happens that $(dxdy)^{2}=-1$.

For spherical symmetry, we would construct, in addition, an operator $K$
defined by $(K+1)u\equiv \sum_{i=1}^{3}\chi _{i}u\vee w_{i}$ and show that $%
\sum_{i=1}^{3}\chi _{i}^{2}u=-K(K+1)u.$ The commutation relations are $\chi
_{i}\chi _{j}-\chi _{j}\chi _{i}=-\chi _{k}$ \ Again, there is no need to
invoke the unit imaginary for anything.

\section{Units imaginary in tangent Clifford algebra}

\subsection{Units imaginary in K\"{a}hler's algebra}

So far, we have dealt with K\"{a}hler's algebra, i.e. Clifford algebra of
differential forms. He writes solutions with rotational symmetry around the $%
z$ axis in configuration space as%
\begin{equation}
u=pe^{im\phi }\tau ^{\pm },\text{ \ \ \ \ }\tau ^{\pm }\equiv \frac{1}{2}%
(1\pm idxdy)=\frac{1}{2}(1\pm iw),
\end{equation}%
where, as he states, $p$ is a genuine meridian differential. A meridian
differential form whose pull-back to the ($\rho ,\phi ,z$) coordinate system
does not depend on $d\rho $ and $dz$. They are said to be genuine (our
translation from the German ``reines'') if, in addition, they do not depend
on ($\rho ,z$). In other words, all the dependence on $\phi $ and $d\phi $
is in $e^{im\phi }\tau ^{\pm }$. It is clear that we can replace $\rho ,$ $z$%
, $d\rho $ and $dz$ with $r,$ $\theta $, $dr$ and $d\theta $ in what has
just been said.

The unit imaginary is not present in (17), but it can be introduced there so
that proper values of angular momentum for (18) be real. K\"{a}hler
introduces the unit imaginary in the exponential, which he did not need
appear in (17). He does not state. We have to assume that the argument is
the standard one in quantum mechanics.

Also, the factor of $dxdy$ has to be of square minus one for $\tau ^{\pm }$
to be idempotents. The algebra can be decomposed into two complementary such
ideals%
\begin{equation}
Cl^{3}=Cl^{3}\frac{1}{2}(1+idxdy)+\frac{1}{2}(1-idxdy),
\end{equation}%
where $Cl^{3}$ denotes in this case the K\"{a}hler in 3-D Euclidean space.
Idempotents, and specially the primitive ones, play a very large role in K%
\"{a}hler's approach to quantum mechanics and, less explicitly so, in the
Dirac theory.

There are different real alternatives ---whose square is minus one--- to the
unit imaginary. They suggest themselves when one uses Clifford algebra
throughout, i.e. for the valuedness of differential forms. K\"{a}hler
considered tensor-valued ones, when not simply scalar-valued. Let us start
by recalling basic features of tangent Clifford algebra.

\subsection{The true tangent Euclidean algebra}

Clifford algebra is the true Euclidean algebra. The usual vector algebra in $%
E^{3}$ is a corruption of Clifford algebra, as it is not available in
dimensions other than three and seven (though these two are of different
nature). Exterior products exist in any number $n$ of dimensions, regardless
of whether the vector space upon which we build the exterior algebra is
Euclidean (i.e. endowed with a vector product) or not. Their combination
with the dot product yields the Clifford product if the vector space is
Euclidean and at least one of the two factors is a vector.

The vector product is a combination of the exterior product and Hodge
duality. The exterior product of two vectors is of grade two, like planes,
with which they are intimately related. Hodge duality assigns to each object
of grade $2$ in a Clifford algebra an object of grade $n-2$ in the same
algebra. \ In $E^{3}$, the duality operation on that exterior product yields
an object of grade one, which is the vector to which we refer to as vector
product. For that reason, a concept of vector product in that sense only
exists in dimension three.

Rotations using Clifford algebra hold the key to understanding our
replacements for the unit imaginary.

\subsection{Rotations in the tangent Clifford algebra}

In a plane, let $\mathbf{u}$ and\ $\mathbf{u}^{\prime }$ be each other's
reflected vector with respect to the unit vector $\mathbf{t}$. Then,
clearly, $\mathbf{tu}^{\prime }=\mathbf{ut}$ where juxtaposition represents
Clifford product. Let $\mathbf{n}$ be a unit normal to $\mathbf{t.}$ By a
theorem of elementary geometry relating reflections with respect to
perpendicular directions, we have%
\begin{equation}
\mathbf{u}^{\prime }=\mathbf{t}^{-1}\mathbf{ut=-n}^{-1}\mathbf{un,}
\end{equation}%
Thus $\mathbf{u}^{\prime }=\mathbf{-n}^{-1}\mathbf{un}$ applies for the
reflection with respect to a plane which has $\mathbf{n}$ as a normal unit
vector.

By another theorem of elementary geometry, the product of reflections with
respect to two planes yields a rotation. Let $\mathbf{n}_{1}$ and $\mathbf{n}%
_{2}$ be the respective normals. The result of rotating $\mathbf{u}$ in this
way yields%
\begin{equation}
\mathbf{u}^{\prime }=\mathbf{n}_{2}^{-1}\mathbf{n}_{1}^{-1}\mathbf{un}_{1}%
\mathbf{n}_{2}=(\mathbf{n}_{1}\mathbf{n}_{2})^{-1}\mathbf{u(n}_{1}\mathbf{n}%
_{2}),
\end{equation}%
the angle $\phi $ being double the angle between the planes, oriented from
the first to the second plane.

Let $\mathbf{N}$ be the unit bivector ($\mathbf{N}^{2}=-1$) in the plane of $%
\mathbf{n}_{1}$ and $\mathbf{n}_{2}.$ We then have%
\begin{equation}
\mathbf{n}_{1}\mathbf{n}_{2}=\mathbf{n}_{1}\cdot \mathbf{n}_{2}+\mathbf{n}%
_{1}\wedge \mathbf{n}_{2}=\cos \frac{\phi }{2}+\sin \left( \frac{\phi }{2}%
\mathbf{N}\right) \mathbf{=}e^{\frac{\phi }{2}\mathbf{N}}.
\end{equation}%
So, finally,%
\begin{equation}
\mathbf{u}^{\prime }=e^{-\frac{\phi }{2}\mathbf{N}}\mathbf{u}e^{\frac{\phi }{%
2}\mathbf{N}}.
\end{equation}

Assume that, instead of tensor-valuedness of differential forms (as K\"{a}%
hler considered), we were interested in valuedness in a tangent Clifford
algebra. Any member of the algebra is a sum of products of vectors. Call any
one such term $\mathbf{U}$ $\mathbf{(=ab...m).}$ Its rotation is given by%
\begin{equation}
\mathbf{U}^{\prime }=(\mathbf{A}^{-1}\mathbf{aA)}(\mathbf{A}^{-1}\mathbf{%
bA)...}(\mathbf{A}^{-1}\mathbf{mA)=}(\mathbf{A}^{-1}\mathbf{ab...mA)=A}^{-1}%
\mathbf{UA,}
\end{equation}%
where $\mathbf{A\equiv e}^{\frac{\phi }{2}\mathbf{N}}.$ And if $\mathbf{U}$
is a member of a left ideal, we have%
\begin{equation}
\mathbf{U}^{\prime }=\mathbf{e}^{-\frac{\phi }{2}\mathbf{N}}\mathbf{U.}
\end{equation}

\subsection{Non-scalar-valued differential forms}

An example of differential form that is not scalar-valued is the
energy-momentum tensor, which is a disguised form of a vector-valued
differential 3-form \cite{V47}. It is a 3-form because it is to be
integrated on a 3-volume in spacetime. It is vector-valued because so is the
result of the integration.

To make the discussion as clear as possible, consider a surface density of a
vector quantity at each point of a surface in 3-D Euclidean space. Let us
denote it as $\mathbf{j}(u,v)du\wedge dv$, where $u$ and $v$ are parameters
on the surface. If we integrate on a very small piece of surface
(appropriately chosen to fit the parametrization), we can approximate the
integral by $\mathbf{j}(u_{0},v_{0})\Delta u\Delta v.$ Its rotation would be
given by $e^{-\frac{\phi }{2}\mathbf{N}}\mathbf{j}(u_{0},v_{0})e^{\frac{\phi 
}{2}\mathbf{N}}\Delta u\Delta v$ where $\mathbf{N}$ is the unit bivector at
the point of the surface with parameters $(u_{0},v_{0})$. It is clear that,
correspondingly, there is the following expression for the rotation of the
vector-valued differential forms:%
\begin{equation}
e^{-\frac{\phi }{2}\mathbf{N}}\mathbf{j}(u,v)e^{\frac{\phi }{2}\mathbf{N}%
}du\wedge dv.
\end{equation}%
The angle of rotation is of course $\phi .$ We should have present the
dissociation of the 3-D vector space where the rotation takes place from the
2-D parameter space. This space is in turn is a proxy for a 2-D submanifold
of the 3-D Euclidean space, which does not change by the rotation. In
addition, one has to express $e^{-\frac{\phi }{2}\mathbf{N}}\mathbf{j}%
(u,v)e^{\frac{\phi }{2}\mathbf{N}}$ in terms of a fixed basis of vectors,
for it does not make any sense to sum components computed on different bases
at different points of the surface.

\subsection{K\"{a}hler's inverse problem and the issue of the unit imaginary
in Lie differentiation and in quantum mechanics}

At least for quantum mechanical physicists, the unit imaginary is
inextricably attached to Lie operators. Such is the case with those for
angular momentum, energy and momentum, which are Lie operators. This is so
not because of Lie theory itself, but because of its use in quantum
mechanics.

In order to understand this, let us give a bird's-eye view of some not yet
considered Lie operator theory in K\"{a}hler's calculus. In 1960, K\"{a}hler
solved what we shall call the inverse problem in Lie differentiation \cite%
{K60}. Given an infinitesimal transformation, obtain a coordinate system in
which the pull-back becomes simply a partial differentiation. It completes
the argument that allows us to view the Lie derivative of a differential
form as a pull-back of a partial derivative. Lie derivatives of differential
transformations $\alpha $ are infinitesimal transformations of $\alpha $,
transformations which in turn are actions of partial derivatives in
appropriate coordinate systems. Of course, all of it up to pull-backs, which
is a redundant statement since that is what a coordinate transformation is
(It is in 1962, not in 1960, that K\"{a}hler referred to the action of an
infinitesimal transformation as a Lie derivative \cite{K62}). Hence, Lie
theory is the relating of partial derivatives of differential forms in
different coordinate systems.

Together with the case we made in our book on differential geometry \cite%
{K60}, differential forms and their derivatives are thus shown to be the
mother and father of all concepts in calculus, analysis and differential
geometry. K\"{a}hler's approach to quantum mechanics is precisely based on
this view, thus on the K\"{a}hler calculus. But, in this calculus, the unit
imaginary has nothing to do with the theory of the Lie derivative. The
introduction of this unit in K\"{a}hler's calculus takes place at a later
stage, the stage of finding differential forms that are solutions of
differential systems with some symmetry property and that also belong to
ideals defined by that property. For those purposes, it can be replaced with
anything os square minus one, like, for example, any of the $w_{i}$'s.

To conclude, the standard unit imaginary has nothing to do with Lie theory
proper, but with the idiosyncrasy of the Dirac equation in context of
solutions that belongs to ideals in a Clifford algebra, and not simply in
the algebra itself.

\section{First geometrization of quantum mechanics}

\subsection{Angular momentum's action on spin idempotents}

K\"{a}hler proposed the form (18) for spatial factors of solutions of
equations with rotational symmetry around the $z$ axis. Although (18) owes
its form to the fact that (17) is what it is, K\"{a}hler does not relate the
two. Had he gone into it, he would have found certain issues with whether
the spin of the electron is given its value of 1/2, if he had analyzed this
issue from the perspective of Eq. (17) (see next subsection). We know that
it is 1/2. But that is in classical treatments, where the electron is,
except for spin, a point particle. But the hydrogen atom treated with the K%
\"{a}hler equation yields exactly the right solutions, and thus includes the
right spin.

In addition for the detailed, exhaustive treatment of the issue of spin,
this author brings to attention from his retrospective perspective that one
must distinguish the arenas for classical systems (or at least classically
treated systems) and quantum systems (or at least quantum-mechanically
treated systems). Both topics will be the subject of future papers.

We shall now limit consideration of the interconnection of (17) and (18) to
the geometrization of the unit imaginary that those equations together
suggest.

\subsection{The unit imaginary in the spin part of angular momentum}

The relation of the spin operator to active rotations is subtle, since, as
we said, the operator acts on differential forms, and rotations act on the
structure where those forms take their value. In addition, one must
distinguish between the form of the rotation of members of the tangent
algebra and of members of the ideals defined by idempotents in the same
algebra that are viewed as not being affected by those rotations.\cite{V44}.

Consider the spin action%
\begin{equation}
\frac{1}{2}w_{3}u-\frac{1}{2}uw_{3},
\end{equation}%
when $u$ is given by (18). If $p$ commutes with $w_{3}$, we have%
\begin{equation}
\frac{1}{2}w_{3}pe^{im\phi }\tau ^{\pm }-\frac{1}{2}pe^{im\phi }\tau ^{\pm
}w_{3}=0.
\end{equation}%
And, if $p$ anticommutes with $w_{3}$,%
\begin{equation}
\frac{1}{2}w_{3}u-\frac{1}{2}uw_{3}=-w_{3}u=-pe^{im\phi }w_{3}\frac{1}{2}%
(1\pm iw_{3}),
\end{equation}%
but%
\begin{equation}
w_{3}(1\pm iw_{3})=w_{3}\mp i=\mp i(1\pm iw_{3}).
\end{equation}%
Hence, proper values are imaginary:%
\begin{equation}
\frac{1}{2}w_{3}u-\frac{1}{2}uw_{3}=\mp iu.
\end{equation}%
The absence of a factor of 1/2 on the right hand side of (31) is the issue
we had in mind in the first paragraph of the previous subsection.

In order to have real proper values, $w_{3}$ in $\chi _{3\text{ }}$ must be
multiplied by some real quantity of square $-1$. From this perspective, $%
\partial /\partial \phi $ (of which (17) is a pull-back) is not real. And
that real quantity must be the same that should replace $i$ in $e^{im\phi }.$
Notice that having a real ``unit imaginary'' in this phase factor is
dictated in last instance by the desire to have a real unit imaginary in the
spin part of the angular momentum operator, not in the orbital part. It was
a result of the interplay of $w_{3}$ and $\tau ^{\pm }$, not $e^{im\phi }.$

\subsection{Geometrization of angular momentum}

In the K\"{a}hler calculus, the issue of the presence or not of the unit
imaginary in the angular momentum operator can then be formulated as
follows. We need, so it seems, (constant) idempotent as factors in solutions
of differential form equations having cylindrical or spherical symmetry.
These idempotents generate ideals of the type $Cl_{3}\tau ^{\pm }$ where $%
Cl_{3}$ is the K\"{a}hler algebra in 3D (i.e. the Clifford algebra of
differential forms for that number of dimensions). The ``unit imaginary'' by
which $\chi _{3\text{ }}$is to be multiplied must be the same as in $\tau
^{\pm }$, which can be expected to be the same as in the rotation operator
in the ($x,y$) plane, i.e. $\mathbf{a}_{1}\mathbf{a}_{2}$, i.e. the $\mathbf{%
N}$ in Eq. (26). Thus, the angular momentum operator should thus be%
\begin{equation}
\chi _{3\text{ }}^{\prime }=\mathbf{a}_{1}\mathbf{a}_{2}\chi _{3\text{ }},
\end{equation}%
corresponding idempotents being%
\begin{equation}
\frac{1}{2}(1\pm dx\mathbf{a}_{1}dy\mathbf{a}_{2}),
\end{equation}%
and (18) becoming%
\begin{equation}
u=pe^{m\phi \mathbf{a}_{1}\mathbf{a}_{2}}\frac{1}{2}(1\pm dx\mathbf{a}_{1}dy%
\mathbf{a}_{2}).
\end{equation}%
$\chi _{3\text{ }}^{\prime }$ so defined is a real quantity. It no longer is
the pull-back of $\partial /\partial \phi $, but of $\mathbf{a}_{1}\mathbf{a}%
_{2}\partial /\partial \phi $. In this way, we have resolved the
antithetical situation about the two meanings of an operator being real. But
the replacement of $\partial /\partial \phi $ with $\mathbf{a}_{1}\mathbf{a}%
_{2}\partial /\partial \phi $ requires a revision of analysis. The pullback
of he orbital part of $\mathbf{a}_{1}\mathbf{a}_{2}\partial /\partial \phi $
to Cartesian coordinates will be written as follows:%
\begin{equation}
\mathbf{a}_{1}\mathbf{a}_{2}\partial /\partial \phi =(x\mathbf{a}_{1})\frac{%
\partial }{\mathbf{a}_{2}\partial y}-(y\mathbf{a}_{2})\frac{\partial }{%
\mathbf{a}_{1}\partial x}.
\end{equation}%
This is a more geometric view of partial derivatives. We leave for other
purposes the development of this approach to analysis, or of calculus if you
will.

\subsection{Other geometrizations of units imaginary}

K\"{a}hler wrote proper functions (meaning proper differential forms)
pertaining to a specific proper value of energy as%
\begin{equation}
u=e^{-iEt}\text{ }\epsilon ^{\pm },\text{ \ \ \ \ \ \ \ }\epsilon ^{\pm
}\equiv \frac{1}{2}(1\mp idt).
\end{equation}%
Following his choice of signature, (-1,1,1,1), we can replace the unit
imaginary with the unit vector $\mathbf{e}_{0}$ in the time direction
(Warning: in order to be consistent with not yet published results obtained
by this author, let us advance that $t$ should be replaced with propertime
viewed as a fifth dimension, and $\mathbf{e}_{0}$ with the unit vector
tangent to the $\tau $ coordinate lines; but we avoid doing so at this time
to avoid confusing readers). Hence, instead of (36), we have%
\begin{equation}
u=pe^{-\mathbf{a}_{0}Et}\frac{1}{2}(1\mp dt\mathbf{a}_{0}),
\end{equation}%
where $p$ depends only on the Cartesian coordinates and their differentials.
The application of the energy operator to this expression does not yield
additional terms since the $d\zeta _{i}$'s are null.

In 1961, K\"{a}hler wrote solutions for given values of proper energy (rest
mass) and proper angular momentum (spin) as%
\begin{equation}
u=e^{im\phi -iEt}p\vee \tau ^{\pm }\vee \epsilon ^{\ast },
\end{equation}%
where, again, $p$ only depends on ($\rho ,z,d\rho ,dz)$ or $(r,\theta
,dr,d\theta ).$ The geometric form of these solutions would be given by%
\begin{equation}
u=e^{m\phi \mathbf{e}_{1}\mathbf{e}_{2}-Et\mathbf{e}_{0}}\vee p\vee \mathbf{%
\tau }^{\pm }\vee \mathbf{\epsilon }^{\ast },
\end{equation}%
with $\tau ^{\pm }$and $\epsilon ^{\pm }$ now replaced with%
\begin{equation}
\mathbf{\epsilon }^{\pm }=\frac{1}{2}(1\mp \mathbf{e}_{0}dt),\text{ \ \ }%
\mathbf{\tau }^{\pm }=\frac{1}{2}(1\pm \mathbf{e}_{1}\mathbf{e}_{2}dxdy),
\end{equation}

Space translation symmetry does not have the same character as time
translation symmetry. No quantum mechanical system has time translational
symmetry. A beam of particles approaches having such symmetry, but as
pertaining to a\ semiclassical treatment where a composite of particles is
treated as if it were a simple system to which the translational symmetry
applied. We may still use the representation%
\begin{equation}
u=pe^{-\mathbf{a}_{1}p_{x}x}\frac{1}{2}(1\pm dx\mathbf{a}_{1}),
\end{equation}%
but we may not expect that all considerations pertaining to time translation
and rotational symmetry will apply to it. The difficulties associated with
position operators in quantum mechanics are well known and need not be
repeated here.

\section{Concluding remarks}

We have simplified K\"{a}hler's demonstating that spin is an overlooked term
in the expression for the pull-back to another coordinate system of a
partial derivative of a differential form. We have then replaced the unit
imaginary with real geometric quantities. This forces one to develop such
geometrization in the future, if those replacements are to be taken
seriously.

As a test for our ideas, consider the hydrogen atom. Replacing $i$ with $%
\mathbf{e}_{0}$ does not alter computations. A harder testing might consist
in replacing the scalar-valued differential forms with Clifford-valued ones
through the replacement $dx^{\mu }$ goes to $dx^{\mu }\mathbf{a}_{\mu }.$
But this hard testing might be spurious, as we proceed to discuss.

In Eq. (34), $dxdy$ is accompanied by $\mathbf{a}_{1}\mathbf{a}_{2}$, but
the other $\mathbf{a}_{1}\mathbf{a}_{2}$ in the same expression is not so
accompanied. Thus, say, $dx$ would not need to go to $dx^{\mu }\mathbf{a}%
_{\mu }$ if it were the result of the product $\mathbf{a}_{1}(dx\mathbf{a}%
_{1})$ in the formulation of a specific K\"{a}hler equation. To complicate
the argument, or rather enrich the realm of possibilities, consider that the
pair of equations by K\"{a}hler and Dirac for the H atom yield the same
result, although the position of the unit imaginary is not equivalent in
those equations.

There are other issues. The concept of a particle in classical physics is
nothing like in quantum physics, where it is not a point but a sophisticated
system, even if the case of an electron. With the benefit of hindsight
provided by his not yet published research, this author wishes to submit
that pure classical and quantum systems will take overlapping sectors of a
Kaluza-Klein type 5-D space. Of course, present day treatments are not pure,
or else the quantum mechanical system of an electron would have already
yielded its mass. As for the appearance of the electron in classical
treatments, we use properties like its mass and its magnetic moment whose
nature is such that one cannot even expect that they will have a classical
formulation in the traditional sense.

The Lorentz transformations belong to the classical sector. Similarities
with this sector and the hybrid nature of the analyses let us obtain results
from this symmetry even in the quantum sector, say, the obtaining of the
fine structure of the hydrogen atom. The Lorentz symmetry is the
manifestation in the spacetime subsector of a symmetry of the geometric
structure, i.e. of the 5-D space. The manifestation in the space-propertime
subspace of the same symmetry is $U(1)\times SU(2)$. As for $SU(3)$, it
would be too complicated to be more specific about it at this point. Suffice
to say that these two symmetries of the quantum sector are as intertwined as
intertwined are $dx$ and $\mathbf{a}_{1}$ in $dx\mathbf{a}_{1};$ $U(1)\times
SU(2)$ and $SU(3)$ belong directly and respectively to the tangent (i.e.
valuedness) algebra and to the K\"{a}hler algebra, i.e. of differential
forms.

\bigskip 

{\LARGE Acknowledgments.} Financial support from PST\ Associates is
acknowledged and deeply appreciated.


\begin{thebibliography}{9}
\bibitem{K62} K\"{a}hler, E.: Der innere Differentialkalk\"{u}l. Rendiconti
di Matematica, 21, 425-523 (1962).

\bibitem{Cartan22} Cartan, \'{E}.: Le\c{c}ons sur les invariants int\'{e}%
graux. Hermann, Paris (1922).

\bibitem{K60} K\"{a}hler, E.: Innerer und \"{a}userer Differentialkalk\"{u}%
l. Abh. Dtsch. Akad. Wiss. Berlin, Kl. Math. Phys. Tech. \textbf{4}, 1-32\
(1960).

\bibitem{V47} Vargas, J.G.: Differential Forms for Cartan-Klein Geometries:
Erlangen Program with Moving Frames, Abramis, London (2012).

\bibitem{V44} Vargas, J. G.: ``The Foundations of Quantum Mechanics and the
Evolution of the Cartan-K\"{a}hler Calculus''. Found. Phys. \textbf{38},
610-647 (1997).

\bibitem{K61} K\"{a}hler, E.: Die Dirac Gleichung. Abh. Dtsch. Akad. Wiss.
Berlin, Kl. Math. Phys. Tech. \textbf{1}, 1-38\ (1961).
\end{thebibliography}
\end{document}